УДК 004.738

# ИНСТИТУЦИОНАЛИЗАЦИЯ ПЛАТЁЖНОЙ СРЕДЫ ЭЛЕКТРОННОЙ КОММЕРЦИИ В РОССИИ
### Калужский М.Л.

***Аннотация:*** Статья о проблемах институционализации платёжной среды электронной коммерции. Автор анализирует тенденции развития платёжных институтов и перспективы создания международного финансового центра в России на основе преимущественного развития электронной коммерции в сферах B2C и C2C. Автор доказывает, что в условиях глобализации рынков потребительских товаров анализ роли и место трансграничных частных платежей в структуре Национальной платежной системы Российской Федерации становятся особенно актуальными.

***Ключевые слова***: электронная коммерция, институциональная теория, сетевая экономика, платёжная среда, платёжный провайдер, электронные деньги, финансовый маркетинг, платёжная система, международный финансовый центр, трансграничные платежи.

# INSTITUTIONALIZATION OF THE PAYMENT ENVIRONMENT OF E-COMMERCE IN RUSSIA
### Kaluzhsky M.L.

***Abstract:*** *Article about problems of Institutionalization of the payment environment of e-commerce. The author analyzes tendencies of development of payment institutes and prospect of creation of the International financial Centre in Russia on the basis of primary development of e-commerce in spheres B2C and C2C. The author proves, that in the conditions of globalization of the consumer markets the analysis of a role and a place of transboundary private payments in structure of National payment system of the Russian Federation become especially current.*

Платёжная среда является важнейшим условием существования электронной коммерции. В сетевой экономике конкурентоспособность её субъектов определяется уровнем развития и взаимной интеграции торговых и платёжных инструментов. Для эффективной коммерческой деятельности необходимо, чтобы банковские учреждения и платежные системы могли «*обрабатывать сделки в режиме реального времени, как внутри страны, так и с использованием зарубежных валют на мировых рынках*» [24, с. 66]. Именно в этом направлении развиваются сегодня платёжные системы и инструменты в сетевой экономике.

**Платежи в электронной коммерции**. В электронной коммерции основным источником институционального развития платёжных систем и инструментов на потребительском рынке является конечный спрос. При этом рост аудитории пользователей онлайн-платежей превышает рост интернет-аудитории в целом, ежегодно увеличиваясь примерно на 3% [18].

Одновременно на рынке электронной коммерции наблюдаются дисбалансы, свидетельствующие об изменении структуры интернет-аудитории и пользовательских предпочтений. Так, например, по данным компании консалтинговой компании «Data Insight», в России по использованию онлайн-платежей лидируют следующие категории населения: молодёжь до 35 лет, люди с высокими доходами, жители г. Москвы и г.

Санкт-Петербурга и опытные пользователи Интернета [18]. Однако в электронной коммерции наблюдается совсем другая тенденция: основной рост покупок приходится на периферийные регионы России, покупателей с невысокими доходами и лиц, недавно подключившихся к Интернету.

В 2012 г. 22 млн. человек покупали товары и услуги в онлайне, что на 30% (5 млн. человек) больше, чем в 2011 году. При этом объем рынка розничной электронной торговли в 2012 году составил около 405 млрд. рублей (~ $ 13 млрд.), из них материальные товары – 280 млрд. руб. Рост по сравнению с предыдущим годом составил 27% [17].

По данным консалтинговой компании «Data Insight» к началу 2013 г. количество пользователей онлайн-платежей в возрасте от 18 до 65 лет, которые когда-либо совершали онлайн-платежи, достигло в России 17 млн. человек (25% аудитории Рунета или 17% населения РФ). При этом прирост пользователей онлайн-платежей стабильно растёт: 3,9 млн. чел. в 2011 г. и 4,3 млн. чел в 2012 г. [18].

Приведённая статистика свидетельствует о том, что если в крупных городах электронная коммерция соседствует с другими причинами пользования Интернетом, то в регионах именно ради онлайн-покупок пользователи осваивают Интернет. При этом поставщики и продавцы товаров в электронной коммерции не менее покупателей заинтересованы в расширении географии интернет-продаж, открывающей перед ними новые рынки. Благодаря онлайн-платежам перед ними открываются совершенно новые горизонты сбыта продукции. «*Web-технологии,* – отмечают И.А. Крымский и К.В. Павлов, – *сделали возможным глобальный электронный финансовый рынок, на котором любое обеспечение, выраженное в любой валюте, теоретически может являться предметом сделки где угодно и когда угодно, а расчеты по этой сделке будут осуществляться немедленно*» [14, с. 80-81].

Электронная коммерция постепенно превращается в один из важных каналов товародвижения, естественным фактором и ограничителем которого являются онлайн-платежи. При том, что спектр электронных платежей чрезвычайно широк. Они осуществляется не только на специализированных сайтах, но и с помощью телефонов, факсимильных аппаратов, компьютерной телефонии, интернет-киосков и банкоматов, платёжных терминалов, интерактивного телевидения и т.д.

Благодаря своей специализации на первое место в организации дистанционных продаж выходят провайдеры платёжных услуг. Базовым субъектом платёжных отношений становится «*процессинговая компания <провайдер>, обеспечивающая программно-техническое взаимодействие между субъектами платежной системы*» [16, с. 75]. Такие компании не просто представляют финансовые услуги, но и обеспечивают функционирование специфического продукта сетевой экономики – «электронных денег».

Законодательство Российской Федерации трактует понятие «электронные деньги» как «*денежные средства, которые предварительно предоставлены одним лицом другому лицу, учитывающему информацию о размере предоставленных денежных средств без открытия банковского счета, для исполнения денежных обязательств … перед третьими лицами*» [23]. Такая трактовка перекликается с определением Европейского парламента и Совета, согласно которому электронные деньги – это «*хранящиеся в электронном виде … денежные средства, представленные в виде требования к эмитенту, которые эмитируются при получении средств для проведения платежных транзакций… и которые принимаются физическим или юридическим лицом, отличным от эмитента электронных денег*» [4, с. 12].

Однако применительно к электронной коммерции больше подходит определение этого понятия, сформулированное К.Г. Миттельман: «*Электронные деньги представляют собой форму денег, выступающую средством осуществления расчетов и отражающую социально-экономические отношения, складывающиеся в рамках виртуальной экономики*» [15, с. 143]. Сегодня существует множество классификаций электрон-ных денег в различной конфигурации и по разным основаниям. [16, с. 90-96] Однако все они имеют общие функциональные особенности:

– сравнительно небольшие суммы платежей;
– целевое предназначение и короткий жизненный цикл;
– недепозитный характер (отсутствие процентов);
– гибкая функциональность (отложенные платежи, платежи с протекцией);
– взаимная независимость провайдеров услуг;
– большое разнообразие клиентов и партнёров.

В той или иной мере электронные деньги используются во всех платёжных системах при расчётах по сделкам в электронной коммерции. Отличия существуют лишь по степени использования электронных денег и характеру электронных платежей. Условно все платёжные услуги можно разделить на два основных вида:

1. *Банковские платежные услуги*, оказываемые населению традиционными банковскими учреждениями. Разнообразие их довольно велико: от пластиковых карт до систем денежных переводов. Денежные переводы доминируют в сфере C2C, тогда как в сфере B2C чаще используются для платы покупок пластиковые карты и банковские переводы.

Переход от традиционных к интернет-трансакциям является всеобщим трендом развития банковской деятельности. Это связано со значительно более низкой стоимостью таких трансакций (см. табл. 1).

Таблица 1. **Относительная стоимость банковских трансакций в США, % [21, с. 26]**

| Вид трансакции | % | |
|---|---|---|
| Личное общение | 100 | * за 100 % принимается стоимость трансакции, осуществленной посредством личного общения |
| Почтовая трансакция | – | |
| Телефонная трансакция | 50 | |
| Интернет-трансакция | 1 | |

Вместе с тем, скорость такого перехода значительно отличается у различных банков. Однако уже сегодня можно выделить две формы оказания банковских услуг: *«услуги, оказываемые так называемыми интернетовскими банками, и услуги, оказываемые традиционными банками, но в онлайновом режиме»* [21, с. 25].

В онлайн-банках *«возникает существенная экономия на обслуживании частных клиентов в результате автоматизации данного процесса…: формирование домашнего банка, создание ЭТП, продвижение платежных схем для электронной торговли и т.п.»* [21, с. 26]. Плюсами являются также отсутствие офисов и неограниченный охват целевой аудитории вне зависимости от местонахождения клиентов.

Наиболее успешным примеров такого подхода в России является банк «Тинькофф Кредитные Системы» («ТКС-Банк»), специализирующийся на обслуживании физических лиц через интернет-сайт, электронную почту и социальные сети. Несмотря на то, что у этого банка нет даже банкоматов, к январю 2013 года им было выпущено свыше 3 миллионов пластиковых карт MasterCard [22]. Однако непосредственно в российской электронной коммерции онлайн-банки пока не играют сколько-нибудь существенной роли.

Безусловным лидером среди традиционных банков, оказывающих связанные с электронной коммерцией платёжные услуги в онлайновом режиме, является «Сбербанк России». В первую очередь это связано с массовым выпуском дебетовых карт «MasterCard Maestro» и «Visa Electron», позволяющих держателям карт получить доступ к безналичным платежам, денежным переводам (комиссия 0-1%) и дистанционному управлению банковским счетом.

Сюда же можно отнести покупку «Сбербанком России» 75% доли в уставном капитале компании «Яндекс.Деньги», крупнейшего в России провайдера электронных платежей. Кроме того, следует отметить, что в июле 2013 г. «Сбербанк-АСТ» (аффили-рованная структура «Сбербанка России») запустила ориентированную на сектор B2C торговую площадку «Сверхмаркет».[1] Если «Сбербанк России» не остановится и пойдёт дальше в направлении интеграции платёжных и торговых инструментов в секторе C2C, то в ближайшие годы он вполне может рассчитывать на роль ведущего провайдера платёжных услуг в России.

Отдельно следует упомянуть о банковских платёжных системах, до недавнего времени доминировавших на рынке платёжных услуг и электронной коммерции: «Contact» («РУССЛАВБАНК»), «Anelik» («Анелик РУ») и др. Они и сегодня достаточно популярны. Падение спроса на их услуги связано с непригодностью таких платёжных систем для мелких сделок (из-за введения минимальной стоимости услуг) и неинтегрируемости с инструментами электронной коммерции.

Несколько особняком стоят платёжные системы, ориентирующиеся на приём платежей через платёжные терминалы «Qiwi» («КИВИ Банк»), «CyberPlat» и ряд др. По оценкам «Эксперта РА», в 2012 году в России более 35% всех платежей физических лиц осуществляются именно с их использованием [7, с. 4].

В целом о потенциале рынка банковских электронных платежей свидетельствует то, что по данным рейтингового агентства «Эксперт РА» совокупный объём банковских онлайн-платежей россиян только в 2012 г. увеличился в 1,5 раза. В этом же году через Интернет физическими лицами совершалось 30% от всех банковских транзакций (прирост на 4%). Тогда как доля интернет-платежей в денежном выражении выросла до 12% (в 2009–2011 гг. их удельный вес не превышал 8-9%). Прогнозируется, что к 2014 г. этот показатель возрастёт не менее чем до 14% [7, с. 4].

2. *Небанковские платежные услуги*, основанные на электронных деньгах и оказываемые специализированными платёжными провайдерами. Они также достаточно разнообразны: от виртуальных платёжных сервисов («Yandex.Деньги», «WebMoney», «RBK Money», «Деньги@Mail.Ru» и др.) до современных процессинговых центров («PayU», «PayOnline» и др.).

Основное преимущество небанковских платёжных услуг заключается в их виртуальности и большей пригодности для приёма мелких платежей. *«Защищённые сделки с кредитными карточками подходят для оплаты крупных сделок, но для продавцов недорогих товаров … система онлайновых платежей (online digital payments, ODP), –* отмечают американские маркетологи Т. Кент и О. Омар, *– … <является> залогом успешных продаж»* [8, с. 695].

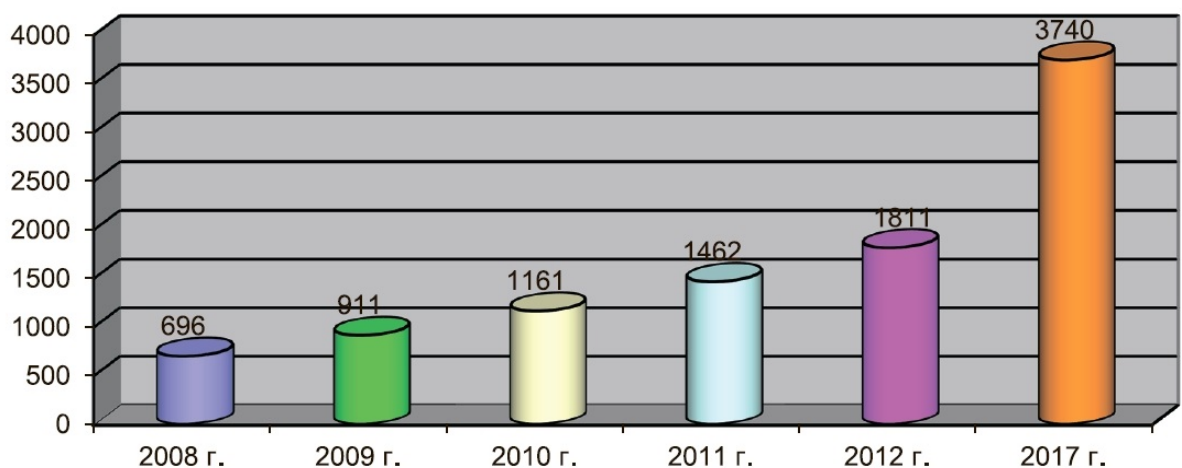

**Рисунок 1. Оборот дистанционных платёжных сервисов, млрд. руб. [5, с. 29]**

Небанковские электронные платёжные услуги в наибольшей степени связаны с электронной коммерцией. У этого много причин: от отсутствия минимального размера платежа до высокой доступности услуг. Именно поэтому они в последнее время переживают значительный рост. Так, например, в 2012 г. оборот дистанционных платёжных сервисов в России достиг 896 млрд. рублей. (49% оборота всех электронных платежей), а к 2017 г. их доля в электронных платежах должна возрасти до 70%, превысив 2,6 трлн. рублей [5, с. 29].

---
[1] Сайт торговой площадки «Сверхмаркет» (ЗАО «Сбербанк-АСТ»). – http://sverhmarket.ru.

На этом рынке сегодня наилучшие показатели роста демонстрируют небанковские сервисы, которые в 2012 г. показали рекордный годовой прирост объёма платежей (72% или 281 млрд. руб.). Пока этот сегмент является самым консолидированным, так как совокупная доля «Visa QIWI Wallet», «Web Money» и «Яндекс.Деньги» составляет около 90% оборота всех электронных денег в России [5, с. 29].

Дальнейшая институционализация платёжных инструментов в электронной коммерции, по всей видимости, будет связана с небанковскими платёжными сервисами. Здесь наблюдается всеобщий для электронной коммерции процесс отказа от использования традиционной инфраструктуры. Как отмечают И.А. Крымский и К.В. Павлов: «*Web-технологии позволяют пользователям финансовых услуг обходиться без посредничества банков – появляются первые признаки дезинтермедиации процесса реализации банковских продуктов*» [14, с. 81].

Отличие банковской сферы от розничной торговли заключается лишь в том, что платёжные операции тут не составляют сколько-нибудь значительного объёма оказываемых услуг. В электронной коммерции революция происходит не столько в сфере банковского, сколько в сфере расчётно-кассового обслуживания субъектов торговли. Поэтому электронные платежи в равной мере являются новой сферой электронного бизнеса, как для банковских, так и для небанковских учреждений. Банковские структуры обладают большими ресурсами и тактически выигрывают в конкурентной борьбе. Тогда как небанковские платёжные провайдеры обладают меньшими трансакционными издержками в отдельно взятой сфере услуг и поэтому имеют стратегические преимущества.

**Регулирование платёжных отношений**. Использование электронных денег в расчётах между участниками электронной коммерции требует изменения регулирующего влияния государства. Это касается не только разработки соответствующего законодательства, но и нормативного закрепления стандартов их использования, права и обязанностей сторон заключаемых сделок. Как отмечают Т. Кент и О. Омар: «*существование электронных денег благотворно отразится на розничной торговле. Однако, чтобы добиться успеха в этой сфере, необходимо разработать чёткие основополагающие правила в отношении стандартов..., создать общий протокол использования таких денег*» [8, с. 696].

На уровне государственного регулирования речь идёт о формировании международных стандартов оказания платёжных услуг, позволяющих превратить их из механизма финансовых взаиморасчётов в фактор обеспечения глобальной конкурентоспособности государства. Не случайно многие зарубежные авторы прямо указывают на то, что «*страны должны рассматривать электронные финансы как способ достигнуть глобальных стандартов, вне использования ресурсов переговорного процесса в рамках ВТО*» [24, с. 66].

Нельзя сказать, что понимание необходимости этой работы в России сегодня отсутствует. Например, министр связи и массовых коммуникаций РФ Н.А.Никифоров в ходе заседания президиума Совета по модернизации прямо указал на то, что необходимо «*разработать ряд национальных технологических стандартов взаимодействия устройств при осуществлении электронных платежей, обеспечить нормативно повсеместность приема платежей в электронной форме и, возможно, с помощью тех же мер государственного регулирования стимулировать наличие технологии для беспроводных платежей (NFC), в том числе, в завозимом в Россию оборудовании*» [3].

Однако сегодня процесс институционализации платёжных инструментов электронной коммерции в России всё же несколько отстаёт от общемировых тенденций. Это связано с многоаспектностью стоящих проблем, затрагивающих интересы большинства министерств и Центрального банка РФ. Эти проблемы требуют пересмотра *«государством отношения к кредитно-денежной системе в целом и банковскому сектору»* [21, с. 23]. Их комплексное решение возможно только на законодательном уровне и на уровне Правительства РФ.

Отчасти таким решением является Федеральный закон Российской Федерации от 27 июня 2011 г. № 161-ФЗ г. «О национальной платежной системе». Этот закон *«устанавливает правовые и организационные основы национальной платежной системы, регулирует порядок оказания платежных услуг, в том числе осуществления перевода денежных средств, использования электронных средств платежа, деятельность субъектов национальной платежной системы, а также определяет требования к организации и функционированию платежных систем...»* [23]. Закон впервые вводит в юридический оборот понятие «платёжная система», определяемая как *«совокупность организаций, взаимодействующих по правилам платежной системы в целях осуществления перевода денежных средств, включающая оператора платежной системы, операторов услуг платежной инфраструктуры и участников платежной системы, из которых как минимум три организации являются операторами по переводу денежных средств»* [23].

Институциональная недостаточность закона «О национальной платежной системе» заключается в его излишней технологичности. На потребительском рынке этот закон не решает задачи формирования институциональной основы электронных платёжных инструментов. Не случайно, в нём нет ни одного упоминания, как об электронной коммерции, так и о деятельности процессинговых компаний, определяющих сегодня тренды развития электронной коммерции.

Проблема деструктивности российской законодательной системы носит структурный характер. Для её решения требуются глубокие институциональные преобразования, затрагивающие основы законодательства. Суть проблемы заключается в том, что российское законодательство основано на романо-германской (континентальной) системе права, ориентированной на приоритетность *«фиксации правовых норм, а не общих принципов допустимого поведения участников рынка»*. В результате нормообразующий принцип «разрешено все, что не запрещено» замещается принципом «запрещено всё, что не разрешено» [11, с. 95].

На уровне Правительства РФ существует понимание актуальности стоящей проблемы. Так, например, в проекте «Концепции создания международного финансового центра в Российской Федерации», разработанном Минэкономразвития РФ и одобренном на заседании Правительства РФ от 05.02.2009 г. прямо говорится о том, что: *«Такое положение существенно затрудняет появление инноваций в финансовой сфере – ... предложение новых видов услуг клиентам, внедрение новых инфраструктурных технологий ... . По ряду направлений нормативно-правовая база финансового рынка в России не полностью сформирована и отстает от практики развитых стран»* [11, с. 96].

В качестве эталона при реформировании финансового законодательства концепцией Минэкономразвития РФ предложена англо-саксонская (прецедентная) система права, основанная на нормообразующем принципе «соблюдай или объясняй» (*principles-based regulation*). Такой (экзогенный, по сути) подход подразумевает установление граничных рамок рыночного поведения вместо жёсткой регламентации. Он несколько усложняет работу регулятора, требуя от него постоянного участия в экономических отношениях.

Вместе с тем, *«такая система является более гибкой и может быстрее адаптироваться к инновациям на финансовых рынках, способствует открытости и конку-*

*ренции*» [11, с. 89]. Не случайно Великобритания является безусловным лидером по распространению электронной коммерции в странах «Большой двадцатки». Объем этого сектора в её экономике в три раза превышает российские показатели. В 2010 г. он достиг 121 млрд. фунтов (около $191,5 млрд.), а доля в ВВП – 8,3%. При этом темпы ежегодного прироста интернет-экономики в Великобритании стабильно держатся на показателе в 11% [6, с. 94].

На основе анализа опыта ведущих экономик мира, успешно использующих электронную коммерцию можно выделить три основных направления институционального регулирования на рынке электронных платежей: [20, с. 39]

1) устранение нормативно-правовых препятствий для развития инноваций;
2) создание конкурентных условий для всех участников рынка;
3) разработка стандартов и усовершенствование инфраструктуры рынка.

Ст. 22 Федерального закона РФ от 27.06.2011 г. № 161-ФЗ г. «О национальной платежной системе» по примеру европейских стран вводит понятие «системно значимой» платёжной системы для участников рынка, осуществляющих переводы «*денежных средств по сделкам, совершенным на организованных торгах*» [23]. Вместе с тем, «*неопределенность относительно прав и обязанностей провайдера и потребителей новой услуги может препятствовать предложению нововведения потенциальными провайдерами и/или мешать его принятию пользователем*» [20, с. 34]. Законодатель не должен останавливаться на введении институциональных норм, не раскрыв их содержание и не определив порядок правоприменения.

Тут вполне можно использовать международную практику. Комитетом по платежным и расчетным системам Банка международных расчетов (Базель, Швейцария, март 2003 г.) группы 10 (G10) ещё в январе 2001 г. были сформулированы «Ключевые принципы для системно значимых платежных систем», внедрённые сегодня в более чем 40 странах мира [9, с. 9]:

I. Система должна иметь детально обоснованную правовую базу во всех применяемых юрисдикциях.

II. Правила и процедуры системы должны давать участникам четкое понимание финансовых рисков, которым они подвергаются из-за участия в ней.

III. Система должна иметь четко определенные процедуры управления рисками, устанавливающими ответственность оператора и участников системы.

IV. Система должна обеспечивать окончательный расчёт в дату валютирования, желательно в течение дня, в крайнем случае – к концу дня.

V. Система с многосторонним неттингом должна обеспечивать своевременное завершение дневных расчетов в случае неплатежеспособности участника.

VI. Используемые для расчетов активы предпочтительно должны быть требованиями к центральному банку, либо нести небольшие кредитные риски и риски ликвидности.

VII. Система должна обеспечивать высокий уровень безопасности и операционной надежности и иметь резервные механизмы завершения обработки платежей в течение одного операционного дня.

В международной экономической практике специалистами выделяется три основных подхода к организации институционального регулирования электронных платежей: европейский, северо-американский и азиатский [15, с. 143]. Каждый из них обладает своей спецификой, обусловленной институциональными особенностями, в которых действуют экономические субъекты сетевой экономики.

1. *Европейский подход* – рассматривает электронные деньги, эмитируемые платёжными операторами, в качестве новой формы денег, требующей особого правового режима для осуществления электронных платежей. К основным мерам европейского правового регулирования рынка электронных платежей относят:

1) нормативное закрепление прав и обязанностей участников электронных расчётов;

3) определение параметров систем безопасности электронных расчётов;

4) нормативное закрепление системы отчётности эмитентов электронных денег перед центральным кредитным банком;

5) обеспечение эмитентами электронных денег их мгновенной ликвидности, т.е. возможности беспрепятственного обмена на валюту государства;

6) возможность установления резервных требований к эмитентам электронных денег.

2. ***Северо-американский подход*** – рассматривает электронные платежи и деньги, эмитируемые платёжными операторами в качестве нового вида платёжных (финансовых) услуг. Наряду с европейским подходом (за исключением, пожалуй, Великобритании), североамериканский подход идёт по пути чёткой регламентации применения правовых норм и инструментов финансового регулирования. Основу этого подхода составляет подробное описание требований к работе финансовых инструментов и рынков (т.н. «*rule-based regulation*»).

3. ***Азиатский (китайский) подход*** – не подразумевает чёткого определения основополагающих понятий и подробной нормативной регламентации платёжных процедур. Вместо этого упор делается на быстрое и эффективное решение ключевых вопросов, связанных с осуществлением электронных платежей. Такой подход позволяет экспортно-ориентированной товаропроизводящей экономике Китая принимать основополагающие решения, не дожидаясь формирования законодателями нормативно-правовой базы. В отличие от предыдущих подходов, отправной точкой здесь является экономическая целесообразность принимаемых решений, а не их соответствие формальным «принципам законодательства».

Задача формирования в России конкурентоспособной платёжной среды электронной коммерции облегчается тем, что «*в настоящее время правовое регулирование выпуска и обращения электронных денег практически отсутствует*» [1, с. 47]. Несколько отстав от партнёров в институциональном развитии, российские регуляторы могут воспользоваться лучшими достижениями различных подходов для формирования уникального облика платёжной инфраструктуры электронной коммерции.

На текущей стадии институционального развития видный российский специалист в области электронной коммерции И.А. Стрелец выделяет три основных аспекта реализации государственной политики в отношении электронных платежей [21, с. 24]:

1. *Ускорение денежного оборота за счёт круглосуточности платежей, исключения посредников и сокращения трансакционных издержек.*

Широкое распространение электронных платёжных инструментов способно вызвать глубокую трансформацию всей платёжной и финансово-банковской инфраструктуры. Многие банки и традиционные платёжные операторы (например, ФГУП «Почта России») уже сегодня сталкиваются с оттоком клиентов в пользу провайдеров электронных услуг.

2. *Потеря дохода Центрального банка РФ от сеньоража*, которую автор предлагает решать с помощью лицензирования эмиссии электронных денег или их самостоятельной эмиссии Центральным банком РФ.

Окончательно эту проблему решить вряд ли удастся. Общая тенденция к сокращению трансакционных издержек в банковском секторе за счёт виртуализации участников рынка и оказываемых ими услуг неминуемо коснётся и доходов Центрального банка РФ. Обнадёживает лишь то, что эти доходы не столь велики и значимы, чтобы ради них ограничивать внедрение инноваций.

3. *Проблема финансового контроля*, усугубляемая усложнением наблюдения за электронными платежами из-за исключения традиционных банковских посредников при проведении электронных трансакций.

Решение проблемы следует искать в трансформации надзорной деятельности самого Центрального банка. Изменение структуры рынка и появление участников, использующих в своей деятельности принципиально новые технологии, действительно делает неактуальным прежние схемы банковского надзора. Однако это не мешает внедрять новые схемы, основанные на использовании новых технологий и инструментов.

Развитие электронной коммерции в России зашло уже столь далеко, что инфраструктура её платёжной среды на потребительском рынке развивается независимо от действий или бездействия финансового регулятора. Ограничительные меры в этой ситуации не приводят к желаемому результату, негативно отражаясь на конкурентных возможностях отечественных товаропроизводителей и развитии платёжной среды в России.

Времена «железного занавеса» прошли, Россия вступила в ВТО, а мы живём в глобальном мире, в котором недостаточная активность одних субъектов с лихвой компенсируется избыточной активностью других. На нашем потребительском рынке отечественные платёжные инструменты легко заменяются услугами транснациональных платёжных сервисов («PayPal», «Skrill», «AliPay» и др.).

Хотя следует отметить, что и традиционные провайдеры обычно находят способ обойти препятствия для платежей в электронной коммерции, устанавливаемые регулятором. Достаточно вспомнить фразу о том, что «денежный перевод не связан с осуществлением предпринимательской деятельности» на бланках платёжной системы «Contact» или анонимные кошельки сервиса «Yandex.Деньги».

Поэтому на уровне государственного регулирования деятельности электронных платёжных систем речь должна идти скорее о содействии в развитии, чем об ограничении деятельности. В качестве примера конструктивной политики можно привести содержание п. 5.2.3. Отчёта «Политика центральных банков в области розничных платежей», подготовленного Комитетом по платежным и расчетным системам Банка международных расчетов (Базель, Швейцария), где говорится: *«Некоторые центральные банки могут там, где это необходимо, дополнить деятельность в качестве катализатора, или органа, содействующего развитию, деятельностью в качестве надзорного органа»* [20, с. 41].

От политики законодателя и финансового регулятора зависит, станет платёжная инфраструктура электронной коммерции фактором инновационного развития и модернизации экономики России или будет задавлена более успешными зарубежными конкурентами. Текущая стадия её институционального развития предоставляет российскому финансовому сектору уникальную возможность не только встроиться в инфраструктуру глобального рынка электронной коммерции, но и занять там лидирующие позиции.

**Перспективы развития платёжной среды**. «Концепцией долгосрочного социально-экономического развития Российской Федерации на период до 2020 г.», принятой Правительством РФ в 2008 году, предусматривается создание в России международного финансового центра (МФЦ) в качестве одного из путей перехода от экспортно-сырьевой к инновационной модели экономического роста. В качестве основной цели макроэкономической политики этим документом определено *«расширение и укрепление внешнеэкономических позиций России, повышение эффективности её участия в мировом разделении труда через поэтапное формирование интегрированного евразийского экономического пространства»* [10].

Предполагается, что к 2020 году *«Россия укрепит свое лидерство в интеграционных процессах на евразийском пространстве, постепенно становясь одним из глобальных центров мирохозяйственных связей (в том числе в качестве международного фи-* *нансового центра) и поддерживая сбалансированные многовекторные экономические отношения с европейскими, азиатскими, американскими и африканскими экономическими партнерами*» [10]. Для реализации поставленных задач необходимо не просто обеспечить создание соответствующей финансовой инфраструктуры, но и освоить пустующие ниши международного рынка финансовых услуг.

При этом нельзя сказать, что у России в этом деле отсутствуют конкуренты: Казахстан сегодня претендует на создание МФЦ, рассчитанного на среднеазиатские государства, а Польша – на государства Центральной и Восточной Европы, а также Украины [11, с. 85]. В Казахстане этот процесс по объективным причинам очень далёк от завершения, а вот Польша уже сегодня по многим показателям заметно опережает Российскую Федерацию [11, с. 85-87].

Слабым местом Польши и Казахстана является стратегия, направленная на создание финансовых институтов копирующих существующие МФЦ, уже лидирующие на глобальном финансовом рынке. Тут трудно не согласиться с авторами одобренной в 2009 г. Правительством РФ «Концепции создания международного финансового центра в РФ», которые отмечают: «*Стратегия догоняющего развития является успешной, пока есть возможность копирования лидера. Однако когда для дальнейшего роста необходимо самому генерировать финансовые инновации, то на первый план выходит уровень развития институтов и человеческого капитала*» [11, с. 93].

Применительно к глобальному маркетингу финансовых услуг у Ф. Котлера есть очень хорошая фраза: «*то, что сработало вчера, скорее всего не сработает сегодня и уж наверняка не сработает завтра*». Это ещё раз говорит о том, что любой эндогенный подход к институциональному планированию неизбежно ведёт к фиаско, так как экономический успех определяется рыночной средой и потребительским спросом. Уже освоенные рыночные ниши контролируются опытными игроками, которые не намерены делиться прибылью и обладают значительно большими ресурсами, чем новые субъекты рынка.

В конкурентной борьбе гораздо важнее то, что глобальная маркетинговая среда финансового рынка «*претерпела значительные изменения за два последних десятилетия, породив новые возможности и новые проблемы*» [13, с. 951]. Поэтому конкурентные преимущества новых МФЦ должны быть связаны с новыми рыночными нишами и связанными с ними возможностями. Только тогда можно будет говорить о благоприятных перспективах их развития.

В России такая ниша существует не в сфере глобальных финансов и фондового рынка, где наша страна не обладает сколько-нибудь значимыми преимуществами, а в сфере трансграничных денежных переводов и электронной коммерции. Правовая база для этого уже создана принятием 22.06.2005 г. «Концепции сотрудничества государств-членов Евразийского экономического сообщества в валютной сфере», в которой формирование единой платежно-расчетной системы связано с перспективой введения единой валюты ЕврАзЭС [12].

Страны ЕврАзЭС не обладают развитой платёжной инфраструктурой, находясь в зависимости от финансовой системы России. Однако товарные поставки на их потребительские рынки идут не из России, а из Китая и других стран. Сложилась ненормальная ситуация, когда Россия фактически выступает донором национальных финансовых систем ЕврАзЭС, так как денежные переводы из России существенно превышают обратные поступления. В 2009 г. эта разница составляла $ 2 млрд. (от $ 8 млрд.), а в 2010 г. она достигла почти $ 2,4 млрд. (от $ 11 млрд.) [2, с. 12].

Де-факто российский рубль уже сегодня выполняет, наряду с долларом США, функцию единой валюты внутри ЕврАзЭС. Так, по итогам 2010 г. на долю России приходилось около 50% трансграничного валютного оборота ЕврАзЭС, из них более 70% – исходящие платежи и только 27% – входящие. Аналогичная пропорция наблюдается и в структуре рублевых переводов, которые составляют примерно треть от всех трансграничных денежных переводов в ЕврАзЭС [2, с. 12].

Большую часть этих денежных переводов из России составляют отправления гастарбайтеров. Однако нельзя забывать о том, что рублёвый рынок и русскоговорящая среда – идеальная среда для продвижения российских товаров народного потребления и технологий. Мы не можем заместить отечественной продукцией поставки товаров из Китая. Но, благодаря электронной коммерции, мы может пустить эти товары через российские товаропроводящие сети, а оплату за них – через российские платёжные системы. Это позволит вернуть часть денежных средств, вывозимых сегодня из Российской Федерации, и заложить институциональную основу для будущего продвижения российской продукции.

Мало того, ничего не мешает повторению в рамках ЕврАзЭС комбинации мировых товаропроизводителей с выводом производственных мощностей в страны Юго-Восточной Азии. При этом у стран ЕврАзЭС, помимо дешевизны рабочей силы, есть большое преимущество, связанное с ограниченностью внутреннего рынка среднеазиатских стран. Поэтому создание там производственных мощностей и передача технологий не приведёт к последующей утрате контроля над ними.

Традиционно низкая покупательная способность в азиатских странах ЕврАзЭС и СНГ не позволяет сегодня использовать для продвижения продукции традиционные каналы сбыта. На потребительском рынке этих стран доминируют устаревшие формы розничной торговли. Однако электронная коммерция снимает все ограничения при условии роста проникновения Интернета.

Причём, ключом для создания единого экономического пространства является формирование международного финансового центра (МФЦ) в рамках ЕврАзЭС, основанного на использовании в трансграничных расчётах как минимум российской платёжной инфраструктуры, и как максимум – российской валюты. Преимущества этого МФЦ будут располагаться не сфере фондового рынка и банковского сектора, а в сфере электронных платежей, выполняемых в интересах товаропроизводителей и частных лиц на потребительском рынке.

Речь идёт не только об использовании российского рубля в трансграничных платежах и денежных переводах, а о формировании на постсоветском пространстве единого экономического пространства, обеспечивающего глубокую инфраструктурную интеграцию его экономических субъектов. Как совершенно справедливо отмечает Ф. Котлер, *«исчезновение барьеров, связанных с необходимостью обмена валют, … позволит расширить межгосударственную торговлю»* [13, с. 953]. Тогда как межгосударственная торговля послужит инструментом взаимной интеграции и экономической модернизации с центром в Российской Федерации.

Для решения указанной задачи потребуется большая подготовительная работа, связанная с глубокими институциональными и инфраструктурными изменениями в национальной платёжной системе самой России. В качестве отправной точки для реформирования платёжных отношений в России можно обратиться к зарубежному опыту. Лидирующие МФЦ обладают вполне определенными конкурентными преимуществами и недостатками, анализ которых может облегчить позиционирование российского МФЦ.

Две модели определяют сегодня тенденции развития платёжной среды электронной коммерции в мировой экономике – американская и китайская. В основе обеих моделей лежит преимущественное развитие платёжных систем логистических уровней 5PL (сетевая логистика) и 4PL (интегрированная логистика). При этом США были первооткрывателями новых платёжных технологий, а Китай творчески их переработал, превратив в инструмент обеспечения глобальной экспансии своих товаров.

Наибольшей перспективой в электронной коммерции обладают платёжные системы уровня 5PL, которые в отличие от уровня 4PL полностью виртуальны и не требуют наличия платёжных терминалов. Распространение платёжных систем уровня 4PL сегодня объясняется недостаточностью проникновения Интернета. С развитием интернет-технологий их значение будет неминуемо сокращаться.

Платёжные провайдеры уровня 5PL столь же виртуальны, как и сама электронная коммерция. Они существуют благодаря и за счёт традиционных провайдеров банковских услуг – карточных платёжных систем, вне которых их клиенты не могут ни снять деньги, ни пополнить свои счета в платёжной системе. Это позволяет платёжным провайдерам легко уходить от множества проблем, связанных с необходимостью соблюдения требований финансовых регуляторов, так как они не эмитируют электронные деньги, а выступают посредниками в платежах.

Базовыми платёжными системами для провайдеров уровня 5PL являются карточные системы «VISA» и «MasterCard», на долю которых приходится сегодня около 50% всех пластиковых карт в мире. Сами провайдеры 5PL представляют собой международные процессинговые центры, глубоко интегрированные в инфраструктуру электронной коммерции, осуществляющие электронные платежи с карточных счетов в реальном времени.

Первым в мире платёжным провайдером уровня 5PL стала американская платёжная система «PayPal». Её институциональная особенность заключается в том, что она изначально была интегрирована в крупнейшую в мире торговую площадку «eBay», являясь её платёжным подразделением. Не случайно доходы от деятельности «PayPal» составили в 2012 году 40% от всех доходов «eBay Inc.».[2]

Уникальный опыт, полученный «PayPal» при проведении платежей на торговой площадке «eBay», был использован для экспансии платёжной системы по всему миру. В результате «PayPal» превратился сегодня в крупнейшего международного провайдера электронных платежей, доходы которого почти на 51% формируются за пределами США. Так, по состоянию на середину 2013 г. «PayPal» работал в 193 странах с 25 национальными валютами и более 132 млн. пользователей. Сегодня платёжная система «PayPal» локализована в 80 странах мира, в том числе и в России. Количество ежесуточных платежей через «PayPal» достигло в 2013 г. (2 квартал) 7,7 млн. трансакций.

Платёжная система «PayPal» из года в год демонстрирует феноменальные показатели роста. В 2012 г. общий оборот сделок через неё достиг $ 145 млрд. (годовой прирост на 22%), из них на долю электронной коммерции пришлось $ 97 млрд. (годовой прирост на 25%). Ежегодный доход «PayPal» достиг в 2012 г. $ 5,6 млрд. (годовой прирост на 26%). При этом более 50% этой суммы пришлось на трансграничные сделки и денежные переводы, а около 25% всех операций «PayPal» (около $ 36 млрд.) – на трансграничную электронную коммерцию.

Альтернативная «PayPal» китайская платёжная система «Alipay» уровня 5PL была создана в 2004 г. для обслуживания платежей на китайских торговых площадках группа компаний «Alibaba Group».[3] Она базируется на использовании пластиковых карт крупнейшей в мире китайской национальной платёжной системы «UnionPay», глобальная эмиссия которой достигла в 2012 г. объёма в 3,4 млрд. пластиковых карт, а также «VISA» и «MasterCard».

Сегодня «Alipay» является крупнейшим платёжным провайдером в Китае, насчитывающим более 800 млн. аккаунтов, увеличившись в 4 раза за период с 2009 по 2012 гг. Кроме «Alibaba Group» эта платёжная система обслуживает ещё более 460.000 тор-говцев по всему миру в 14 основных иностранных валютах. В России «Alipay» обслуживает платежи через платёжные системы «QIWI», «WebMoney», не считая платежей с банковских карт «VISA» и «MasterCard».

---

[2] Сайт платёжной системы «PayPal». – https://www.paypal-media.com/about

[3] Группа компаний «Alibaba Group». – http://news.alibaba.com/specials/aboutalibaba/aligroup/index.html

Общей особенностью платёжных провайдеров «PayPal» и «Alipay» является выполнение функций институционального регулирования и арбитража в электронной коммерции. Так, «PayPal» принимает и рассматривает в течение 45 дней с момента покупки претензии покупателей, а также списывает со счёта продавцов в безакцептном порядке денежные средства в случае принятия решения в их пользу. «Alipay» делает то же самое, резервируя оплату до истечения контрольного срока с момента получения покупки (по данным транспортного провайдера). В случае если клиент недоволен покупкой, «Alipay» полностью возвращает произведённую покупателем оплату.

Феноменальный успех платёжных провайдеров уровня 5PL на потребительском рынке объясняется их глубокой интеграцией в торговую инфраструктуру платёжной коммерции. Покупатели ничем не рискует, приобретая товар – он всегда может получить деньги обратно, если товар не соответствует описанию или испорчен при пересылке. Сама возможность мгновенной оплаты товара через Интернет на уровне 5PL формирует высшую степень доверия покупателей к продавцам. Платёжные провайдеры низших уровней (2PL-4PL) не могут ничего противопоставить такой политике, стремительно утрачивая конкурентоспособность.

В России также существуют платёжные провайдеры уровня 5PL способные составить конкуренцию крупным зарубежным платёжным провайдерам. Однако из-за бездействия регулятора они до сих пор не в состоянии полноценно оказывать платёжные услуги в сферах B2C и C2C. Например, российский процессинговый центр «PayU», интегрированный в крупнейшую торговую площадку «Molotok.ru», с 2005 г. успешно работает в Восточной Европе, являясь лидером рынка услуг процессинга в Чехии, Венгрии, Румынии, Польше и ЮАР.[4] И только в России его деятельность не находит должной институциональной поддержки.

Это не означает, что на российском рынке электронной коммерции отсутствуют такого рода платёжные услуги. Сегодня их оказывают ведущие зарубежные провайдеры «PayPal» и «AliPay», весьма успешно ведущие свой бизнес в России. Покупатели также не испытывают больших неудобств. Единственная проблема заключается в том, что электронная коммерция в России развивается по догоняющему пути развития и не является интеграционным механизмом на постсоветском пространстве.

Главный финансовый регулятор – Центральный банк РФ – пока не воспринимает платёжных провайдеров уровня 5PL, оказывающих услуги процессинга платежей в электронной коммерции, в качестве сколько-нибудь значимых субъектов рынка. Так, например, в отчете, подготовленном регулятором в сентябре 2011 года для Комитета по платежным и расчетным системам Банка международных расчетов (Базель, Швейцария), о них даже не упоминается [19].

Вместе с тем, в «Концепции долгосрочного социально-экономического развития Российской Федерации на период до 2020 г.» создание условий для развития компаний, работающих в области электронной торговли, признано Правительством РФ одним из приоритетных направлений развития информационно-коммуникационных технологий в России [10]. Задача состоит в том, чтобы ускорить формирование благоприятной институциональной среды для этих компаний, дав им возможность в полной мере раскрыть инновационный потенциал отечественной электронной коммерции.

---

[4] Сайт ООО НКО "ПэйЮ". – http://www.payu.ru/o-nas.

***Опубликовано***: Калужский М.Л. Институционализация платежной среды электронной коммерции в России // Финансовая аналитика: проблемы и решения. – 2014. – № 2 (188). – С. 8-19. – ISSN 2073-4484. (Доступна электронная версия).